\documentclass[twocolumn]{revtex4}
\usepackage{amssymb,epsf}
\usepackage{latexsym}
\usepackage{xcolor}
\usepackage{epsfig}
\usepackage{float}
\usepackage{amsmath}
\usepackage{graphicx}

\begin{document}
\title{Observational constraints on New Tsallis holographic energy in Rastall theory}

\author{N. Sadeghnezhad$^1$\footnote{nsadegh@maragheh.ac.ir}, R. Jalalzadeh$^1$\footnote{r.jalalzadeh@riaam.ac.ir}, Z. Davari$^2$\footnote{zahradavari@kias.kr.re}, B. Afshar$^1$\footnote{behnoush.afshar.cosmology@gmail.com}}
\address{$^1$ Research Institute for Astronomy and Astrophysics of Maragha
(RIAAM), University of Maragheh, P.O. Box 55136-553, Maragheh,
Iran\\ $^2$ School of Physics, Korea Institute for Advanced Study (KIAS), 85 Hoegiro, Dongdaemun-gu, Seoul, 02455,  Korea}

\today
\begin{abstract}
The cosmological implications of New Tsallis holographic dark energy (NTHDE) in Rastall theory have been studied. Using the data set that includes DESI BAO (DR2), PantheonPlus SNe Ia, $H(z)$ measurements, and BBN and the MCMC analysis, the key cosmological and model-specific parameters are constrained. The result is compared with that of the $\Lambda$CDM model indicating that in addition to providing a viable dynamical dark energy framework, predictions for $H(z)$ are slightly more consistent with intermediate-redshift observations. Generally, the model remains compatible with current data and offers testable deviations from $\Lambda$CDM for upcoming surveys. It is also seen that when the energy density of quantum fields in vacuum, exposed by NTHDE, is combined with the Rastall correction term to the general relativity, a plausible candidate for dynamical dark energy is obtained that mimic the current value of the dark energy density parameter reported in the $\Lambda$CDM model. The latter cannot be repeated by NTHDE alone. The study also confirms previous theoretical and observational constraints on the Rastall parameter obtained by focusing on the thermodynamics, early universe, pulsars, and the early-type galaxies.
\end{abstract}

\maketitle

\section{Introduction}

Generalized statistical theories are formulated to study systems, including long-range interactions
\cite{Renyi,tsallis,pla}. Therefore, interest in investigating gravitational systems such as black holes has been raised
\cite{Tsallis:2012js,non23,non22,5,non18,Czinner:2017tjq,Ghaffari:2019itm},
they reconcile the third law of thermodynamics and black hole physics,
and pave the way to increase our understanding of energy in
gravitational systems \cite{Moradpour:2021soz}. In astrophysics, generalized statistics leads to considerable outcomes in Stellar
models \cite{plas}, Solar neutrinos \cite{Kaniadakis:1996ja},
Stellar formation,, and Bok globules (Jeans mass)
\cite{TKG,Moradpour:2019wpj,Moradpour:2024azo,Moradpour:2024uqa},
and provides justifications for the MOND theory
\cite{Moradpour:2017fmq,Sheykhi:2019bsh}. The corresponding
Cosmological models have also attracted attention to themselves
\cite{Abreu:2012msk,Barboza:2014yfe,Moradpour:2016rcy,Nunes:2015xsa,non19,Moradpour:2017ycq,Sheykhi:2022gzb,Komatsu:2024lnm}
and, for example, a primary inflationary universe without an inflaton field seems possible \cite{ghaffari2020inflation}.
Therefore, deviations from the standard Friedmann equations are not confined to entropy production \cite{Eling:2006aw} and can be expected as the result of employing the generalized statistics in a thermodynamic equilibrium state \cite{Moradpour:2016rcy}. On the other hand, the usage of generalized statistics leads to generalized versions of the Heisenberg uncertainty principle and uncertainty relationships that are also obtained in quantum scenarios of gravity, a result addressing a deep connection between quantum aspects of gravity and generalized statistics
\cite{Moradpour:2019yiq,Moradpour:2020kss}. Indeed, as a non-zero minimum length and a non-zero minimal uncertainty in momentum (quantum gravity scenarios) may solve the $H_0$ tension
\cite{Aghababaei:2021gxe,Moradpour:2022oxr,Nozari:2024wir,CosmoVerse:2025txj},
the generalized statistics-based uncertainty principles and
Relationships also suggest solutions to the $H_0$ controversy
\cite{Sadeghnezhad:2025lxn}.

Reconciling quantum and gravity through modifying the upper bound
of gravitational systems \cite{Cohen:1998zx} leads to a pioneering
hypothesis for relating dark energy to the energy of the quantum
fields in vacuum \cite{Li:2004rb}. It is called holographic dark energy (HDE), and this theory is in agreement with the
thermodynamic laws \cite{Moradpour:2023ayk}. Motivated by the
weaknesses of the original HDE model introduced in
\cite{Li:2004rb} and also the successes of the generalized
statistics in studying astrophysical, cosmological, and black hole
systems, some authors propose to use the generalized statistics'
entropies of gravitational systems to reconcile the energy of quantum
fields in vacuum and dark energy. Correspondingly, new holographic dark energy models were born, which include Tsallis holographic
dark energy (THDE) \cite{Tavayef:2018xwx}, Sharma-Mital
holographic dark energy (SMHDE) \cite{SayahianJahromi:2018irq},
R\'enyi holographic dark energy (RHDE) \cite{Moradpour:2018ivi},
Kaniadakis holographic dark energy (KHDE)
\cite{Moradpour:2020dfm}, and new Tsallis holographic dark energy
(NTHDE) \cite{Moradpour:2020dfm} are introduced. Unlike the
original HDE \cite{Li:2004rb}, all of these models are capable to
derive a suitable model for the universe by employing the apparent
horizon as the IR cutoff, which is important as it is the causal boundary for the cosmos
\cite{Hayward:1997jp,Hayward:1998ee,Bak:1999hd,Cai:2005ra,Cai:2006rs,Akbar:2006kj,Cai:2008gw}.

On the one hand, by employing diverse theories of gravity, various properties of THDE, SMHDE, RHDE, KHDE, and NTHDE have been studied
(see
Refs.~\cite{Tavayef:2018xwx,SayahianJahromi:2018irq,Moradpour:2018ivi,Moradpour:2020dfm,Sharma:2021zjx,Drepanou:2021jiv,Pandey:2021fvr}
and their citations). On the other hand, the new Tsallis entropy
and the corresponding statistics generate considerable results in
various astrophysical and cosmological systems
\cite{Abbasi:2020poo,Moradpour:2020kss,Moradpour:2024azo,Sadeghnezhad:2025lxn}.
Moreover, the cosmological implications of the holographic dark energy models in the non-divergence-less gravitational theories, such as Rastall theory, have also attracted attention \cite{Ghaffari:2020nnk,Maity:2020wqo,Saleem:2021iju,Saleem:2023wyn,Waheed:2023flg,Saleem:2024bhi,Fazlollahi:2024lrt,Fazlollahi:2024mra}.
Rastall theory is a conservative theory in the sense that the energy-momentum tensor is not divergence-less, and in fact, it is in relationship with the divergence of the Ricci scalar \cite{rastall,Smal,Smal1}. Although its Lagrangian is still under discussion \cite{Josset:2016vrq,Dzhunushaliev:2016rdz,Moradpour:2018gkm,Shabani:2020wja,Fabris:2020uey,Smal,Smal1,Licata:2017rfx}, its successes in satisfying diverse observations \cite{Moradpour:2017ycq,Li:2019jkv,ElHanafy:2022kjl,Afshar:2023uyw,Darabi:2017coc} and theoretical expectations (see also citations of Refs.~\cite{Darabi:2017coc,Visser:2017gpz}) motivate us to study its various aspects.

Here, taking into account the Rastall theory, we are going to
study the evolution of a flat FRW universe where the energy of quantum fields in the vacuum are scaled using the NTHDE hypothesis \cite{Moradpour:2020dfm}. There is also no interaction between the cosmic fluids, including NTHDE and a pressureless source as dark matter (DM). Indeed, it is also shown that the combination of the Rastall term and the energy of quantum fields in vacuum (NTHDE) may mimic the predictions of $\Lambda$CDM model about the dark energy density parameter. The parallel goal is also to investigate the allowable range for the Rastall parameter in accordance with the current accelerating universe. In our setup, the apparent horizon is considered as the IR cutoff since it is the proper causal boundary for the FRW universe \cite{Hayward:1997jp,Hayward:1998ee,Bak:1999hd,Cai:2005ra,Cai:2006rs,Akbar:2006kj,Cai:2008gw} and we employ the unit of $c=\hbar=G=K_B=1$.

\section{Tsallis entropy of Black hole and NTHDE}

In addition to the power-law Tsallis entropy of a black hole
\cite{Tsallis:2012js}, which leads to the Tsallis holographic dark energy (THDE) model \cite{Tavayef:2018xwx},
an alternative formulation of Tsallis entropy has also been proposed \cite{pla,Tsallis:2012js}. It is defined as
\begin{eqnarray}\label{1}
S_q^T=\frac{1}{1-q}\sum_{i=1}^{W}(P_i^q-P_i),
\end{eqnarray} where $W$ denotes the total number of accessible microstates of the system and $q$ is the Tsallis non-additivity parameter \cite{pla,Tsallis:2012js}. Here,  $P_i$  represents the probability
of the system occupying the $ i$-th state. The  standard Boltzmann-Gibbs entropy
is recovered in the limit $q\rightarrow 1$, and for $P_i=\frac{1}{W}$, one obtains:
\begin{equation}
 S_q^T=\frac{W^{1-q}-1}{1-q}, 
\end{equation}
consistent with the original Tsallis formulation \cite{pla,Tsallis:2012js}.
Considering a black hole with event-horizon surface area $A$,the corresponding entropy can be expressed as \cite{Moradpour:2020dfm}
\begin{eqnarray}\label{2}
S_q^T=\frac{1}{1-q}[\exp\big(\frac{(1-q)A}{4}\big)-1].
\end{eqnarray}
 The spacetime geometry of a non-flat Friedmann–Robertson–Walker (FRW) universe is described by the metric
\begin{eqnarray}\label{3}
ds^{2} = -dt^{2}+a^{2}\Big(\frac{dr^{2}}{1-K r^{2}}+
r^{2}d\Omega^{2}\Big),
\end{eqnarray}

\noindent where $a(t)$ is the scale factor, and  $K=-1,0,1$ corresponds to open, flat, and closed spatial geometries, respectively. The apparent horizon of the FRW universe is located at
\begin{eqnarray}\label{4}
\tilde{r}_A=\frac{1}{\sqrt{\frac{K}{a^2}+H^2}},
\end{eqnarray}

\noindent and therefore, following the NTHDE proposal introduced in
Ref.~\cite{Moradpour:2020dfm}, the corresponding energy density of quantum fields in vacuum can be written as

\begin{equation}\label{5}
\rho_{\nu}=\frac{3C^2}{8\pi\delta
\tilde{r}_A^4}[\exp\big(\frac{\delta\pi}{\frac{K}{a^2}+H^2}\big)-1],
\end{equation}

\noindent where $\frac{3C^2}{8\pi}$ s an undetermined constant \cite{Li:2004rb,Moradpour:2020dfm}, and 
$\delta\equiv1-q$.
For a flat universe ($K=0$), indicated by WMAP observations
\cite{roos}, we have \cite{Moradpour:2020dfm}

\begin{eqnarray}\label{51}
\rho_{\nu}=\frac{3C^2H^4}{8\pi\delta}[\exp\big(\frac{\delta\pi}{H^2}\big)-1],
\end{eqnarray}

\noindent and thus \cite{Pandey:2021fvr}

\begin{eqnarray}
\rho_{\nu}\approx
\frac{3C^2}{8\pi}\big[\pi
H^2+\frac{\delta\pi^2}{2}+\frac{\delta^2\pi^3}{6H^2}\big],
\end{eqnarray}

\noindent which clearly recovers the Bekenstein entropy result at
the appropriate limit $\delta\rightarrow0$ \cite{Li:2004rb}. It has been shown that in some frameworks such as general relativity, $\rho_{\nu}$ plays the role of dark energy \cite{Moradpour:2020dfm,Pandey:2021fvr}. Our analysis in the following sections reveals that $\rho_{\nu}$ $i$) can accelerate the Cosmos in the Rastall framework and $ii$) when it is combined with the Rastall term, it can mimic the current value of dark energy density parameter predicted in the $\Lambda$CDM model.
\section{Cosmological model}
The gravitational field equations in Rastall theory are given by
\cite{rastall}
\begin{eqnarray}\label{r1}
G_{\mu \nu}+\kappa\lambda g_{\mu \nu}R=\kappa T_{\mu \nu},
\end{eqnarray}
\noindent where $\kappa$ and $\lambda$ represent the Rastall gravitational coupling constant and the Rastall parameter, respectively. The Newtonian limit constrains them as
\cite{rastall,Moradpour:2016ubd}

\begin{eqnarray}\label{kappa}
\frac{\kappa}{4\kappa\lambda-1}(3\kappa\lambda-\frac{1}{2})=\kappa_N,
\end{eqnarray}

\noindent where $\kappa_N=4\pi$ denotes the Newtonian
gravitational coupling constant. Obviously, the choice $\lambda=0$ leads to the well-known general relativity, for which ($\kappa=2\kappa_N$).
For the Friedmann equations and conservation law, one gets
\cite{Moradpour:2015ymo}

\begin{eqnarray}\label{friedman1}
&&(12\kappa\lambda-3)H^2+6\kappa\lambda \dot{H}=-\kappa\rho,\nonumber\\
&&(12\kappa\lambda-3)H^2+(6\kappa\lambda-2) \dot{H}=\kappa p,
\end{eqnarray}

\noindent and

\begin{equation}\label{cont}
(\frac{3\kappa\lambda-1}{4\kappa\lambda-1})\dot{\rho}+(\frac{3\kappa\lambda}{4\kappa\lambda-1})\dot{p}+3H(\rho+p)=0,
\end{equation}

\noindent respectively. The Newtonian limit implies the
existence of only one free parameter, motivating us to define
$\lambda\kappa=\gamma$ \cite{Moradpour:2016ubd} that finally brings the Friedmann equations corresponding to NTHDE model \cite{Moradpour:2020dfm} to the form

\begin{eqnarray}\label{friedman2}
&&H^2+\frac{2\gamma}{4\gamma-1}\dot{H}=\frac{8\pi}{3(1-6\gamma)}\left(\rho_r+\rho_m+\rho_{\nu}\right),\\
&&H^2+\frac{2(1-3\gamma)}{3(1-4\gamma)}\dot{H}=-\frac{8\pi}{3(1-6\gamma)}(p_{r}+p_{\nu}).\nonumber
\end{eqnarray}

\noindent Here, the subscripts  $m$, $r$ and $\nu$ represent the pressure-less dark matter, radiation, and vacuum energy components, respectively. Therefore, one can easily recast the first equation to

\begin{eqnarray}\label{7}
1=\Omega_m+\Omega_r+\Omega_D,
\end{eqnarray}

\noindent by defining

\begin{eqnarray}\label{8}
&&\Omega_{m}\equiv\frac{\rho_{m}}{(1-6\gamma)\rho_c}, \\&&\Omega_v\equiv\frac{\rho_{v}}{(1-6\gamma)\rho_c},\nonumber\\
&&\Omega_{R}\equiv\frac{\frac{6\gamma}{4\gamma-1}\dot{H}}{8\pi\rho_c}=\frac{2\gamma\dot{H}}{(4\gamma-1)H^{2}},\nonumber\\
&&\Omega_{r}\equiv\frac{\rho_{r}}{(1-6\gamma)\rho_c},\nonumber\\
&&\Omega_{D}\equiv\Omega_v-\Omega_R,\nonumber
\end{eqnarray}

\noindent where $\rho_c\equiv\frac{3H^2}{8\pi}$. In the absence of
any interaction between cosmic fluids, Eq.~(\ref{cont}) should be valid for each fluids separately leading to

\begin{eqnarray}\label{09}
\rho_m&=& \rho_{0m}
a^{\frac{-3(4\gamma-1)}{3\gamma-1}}=\rho_{0m}(1+z)^{\frac{3(4\gamma-1)}{3\gamma-1}},\\ \rho_r&=& \rho_{0r} a^{-4},
\end{eqnarray}

\noindent where $z$ is the redshift ($1+z=\frac{1}{a}$). The evolution equation for $\rho_\nu$ becomes

\begin{eqnarray}\label{10}
&&\big[(\frac{3\gamma}{4\gamma-1})\dot{\omega}_{v}+3H(1+\omega_{v})\big]\rho_{v}\nonumber\\
&&+(\frac{3\gamma(1+\omega_{v})-1}{4\gamma-1})\dot\rho_{v}=0.
\end{eqnarray}

\noindent Here, $\omega_\nu\equiv\frac{p_\nu}{\rho_\nu}$ and $\rho_{0i}$ denotes
the current value ($a=1$) of $\rho_{i}$. Hence, employing
Eq.~(\ref{8}), the deceleration parameter is obtained as

\begin{eqnarray}\label{9}
q &=& -1-\frac{\dot{H}}{H^2}=-1 - \frac{4\gamma-1}{2\gamma} \, \Omega_R,
\end{eqnarray}

\noindent where the Raychaudhuri equation

\begin{equation}\label{rey}
\dot{H}=-4\pi\frac{(4\gamma-1)}{(6\gamma-1)}(\rho_m+\rho_{v}+p_{v}+\rho_r+p_r),
\end{equation}

\noindent has been used to derive the second line. It should be
noted that thermodynamic analysis indicates
$\frac{1}{6}<\gamma<\frac{1}{4}$ is not allowed
\cite{Moradpour:2016fur}, a result also confirmed by various
observational studies
\cite{Li:2019jkv,ElHanafy:2022kjl,Afshar:2023uyw}. Moreover, the  redshift-scale factor relationship implies
$\frac{d}{dt}=-H(1+z)\frac{d}{dz}$ used in future analyses, and the Friedmann second equation leads to

\begin{equation}\label{18}
\frac{1-3\gamma}{3\gamma}\Omega_R-1=\frac{(\omega_{v}\Omega_{v}+\frac{1}{3}\Omega_r)}{1-6\gamma},
\end{equation}

\noindent combined with Eq.~(\ref{7}) to finally reach

\begin{equation}\label{19}
\frac{1 - 6\gamma}{1 - 3\gamma} = \Omega_{v}\left(1 + \frac{3\gamma}{1 - 3\gamma}\omega_{v}\right) 
+ \Omega_{m}+\frac{(1 - 2\gamma)}{(1 - 3\gamma)}\Omega_{r},
\end{equation}

\noindent that, just like Eq.~(\ref{7}), clearly covers the
Friedmann's first equation obtained in the framework of general
relativity when $\gamma=0$ \cite{Moradpour:2020dfm}.

\section{Observational constraints and stability}
For our observational datasets, we incorporate baryon acoustic oscillation (BAO) measurements from the DESI 2025 Data Release~\cite{DESI2025}, type Ia supernova luminosity distance data from the PantheonPlus compilation~\cite{Scolnic:2022}, direct Hubble parameter measurements $H(z)$ from cosmic chronometers~\cite{Farooq:2017}, as well as the Big Bang Nucleosynthesis.

To compare the theoretical predictions of our cosmological model with the observations, we directly implement the modified background evolution equations into a numerical solver for the Hubble parameter $H(z)$. We then compute the corresponding observables for each dataset, including supernovae distance moduli, BAO distance ratios, and Hubble parameter values.

For parameter estimation, we perform a Metropolis--Hastings Markov Chain Monte Carlo (MCMC) analysis with multiple chains, each initialized at a different starting point. Convergence is assessed using the Gelman--Rubin statistic $R-1$, with convergence defined as $R-1 < 0.01$~\cite{Gelman:1992zz}. Posterior distributions and confidence contours are generated using the \texttt{GetDist} package.


\subsection{DESI BAO}

Baryonic acoustic oscillations (BAO) are the imprints of sound waves in the photon-baryon plasma of the early Universe. As shown in Table~\ref{tab:desi2025-compressed}, we employ the recently released DESI Data Release~2 (DR2, 2025) BAO dataset~\cite{DESI2025}, 
which provides 13 compressed measurements spanning the redshift range $0.3 \leq z \leq 2.33$. 
Specifically, the dataset consists of six measurements of the comoving angular diameter distance, 
\begin{equation}
    D_M(z) = \int_0^z \frac{c \, dz'}{H(z')} ,
\end{equation}
Six measurements of the Hubble distance,
\begin{equation}
    D_H(z) = \frac{c}{H(z)} ,
\end{equation}
and one isotropic measurement of the volume-averaged distance, 
\begin{equation}
    D_V(z) = \left[ z \, D_M^2(z) \, D_H(z) \right]^{1/3},
\end{equation}
obtained from the Bright Galaxy Survey (BGS) at $z_{\rm eff}=0.295$.

All measurements are reported in dimensionless form $D_M/r_d$, $D_H/r_d$, and $D_V/r_d$. Here $r_d$, the drag-epoch sound horizon, denotes the maximum comoving distance traversed by acoustic waves before baryon-photon decoupling.\\
%
The corresponding $\chi^2_{\mathrm{DESI}}$ function is constructed as
\begin{equation}
\chi^2_{\mathrm{DESI}} = 
\Delta {D}^T \, {C}^{-1}_{\mathrm{DESI}} \, \Delta {D} ,
\end{equation}
where $\Delta {D}$ denotes the difference between the observed and theoretical vectors of $\{ D_M/r_d, D_H/r_d, D_V/r_d \}$, and ${C}^{-1}_{\mathrm{DESI}}$ is the  inverse of the  full $13\times13$ covariance matrix provided in the DESI DR2 release.
The DESI covariance matrix  is defined as:
\begin{equation}
\text{C}_{\mathrm{DESI}} =
\left(
\begin{smallmatrix}
\sigma_1^2 & \rho_{1,2}\,\sigma_1\sigma_2 & \cdots & \rho_{1,13}\,\sigma_1\sigma_{13} \\
\rho_{2,1}\,\sigma_2\sigma_1 & \sigma_2^2 & \cdots & \rho_{2,13}\,\sigma_2\sigma_{13} \\
\vdots & \vdots & \ddots & \vdots \\
\rho_{13,1}\,\sigma_{13}\sigma_1 & \rho_{13,2}\,\sigma_{13}\sigma_2 & \cdots & \sigma_{13}^2
\end{smallmatrix}
\right),
\end{equation}

where $\sigma_i$ are the measurement uncertainties reported (from Table~\ref{tab:desi2025-compressed}) and $\rho_{i,j}$ are the correlation coefficients between the measurements $i$ and $j$.

\begin{table}[ht]
\centering
\caption{BAO measurements from DESI BAO (DR2) used in this work~\cite{DESI2025}}
\label{tab:desi2025-compressed}
\begin{tabular}{|l|c|c|c|c|}
\hline
Tracer & $z_{\mathrm{eff}}$ & $D_M/r_d$ & $D_H/r_d$&$D_V/r_d$  \\
\hline
BGS        & 0.295 & --              & --   &    $ 7.94 \pm 0.08$       \\
\hline
LRG1       & 0.510 & $13.59 \pm 0.17$ & $21.86 \pm 0.43$&-- \\
\hline
LRG2       & 0.706 & $17.35 \pm 0.18$ & $19.46 \pm 0.33$&--  \\
\hline
LRG3+ELG1  & 0.934 & $21.58 \pm 0.15$ & $17.64 \pm 0.19$&--  \\
\hline
ELG2       & 1.321 & $27.60 \pm 0.32$ & $14.18 \pm 0.22$&--  \\
\hline
QSO        & 1.484 & $30.51 \pm 0.76$ & $12.82 \pm 0.52$&--  \\
\hline
Ly$\alpha$ & 2.330 & $38.99 \pm 0.53$ & $8.63 \pm 0.10$&--   \\
\hline
\end{tabular}
\end{table}

\subsection{Type Ia Supernovae (SNIa)}
Type Ia supernovae (SNIa) act as standardizable candles, 
allowing precise distance estimation through their apparent 
brightness and redshift. In this work, we employ the 
Pantheon+SH0ES compilation \cite{Scolnic:2022}, which 
represents the most recent and comprehensive collection 
of spectroscopically confirmed SNIa. 
The Pantheon sample consists of 1701 light curves 
spanning the redshift range $0 < z < 2.3$. Among these, 
77 SNIa are hosted by Cepheid-calibrated galaxies in the 
low-redshift interval $0.00122 \leq z \leq 0.01682$, 
providing a local calibration that is central to the SH0ES 
distance-ladder measurement of $H_0$. The remaining 
Supernovae come from 18 distinct surveys, covering both 
low and high redshift regimes, thereby enabling a precise 
reconstruction of the cosmic expansion history.

The theoretical distance modulus is computed as
\begin{equation}
    \mu_{\rm th}(z) = 5 \log_{10} \left( \frac{D_L(z)}{\,{\rm Mpc}} \right) + 25 ,
\end{equation}
where the luminosity distance is given by
\begin{equation}
    D_L(z) = (1+z) \, c \int_0^z \frac{dz'}{H(z')} .
\end{equation}

The observed magnitude of a supernova is modeled as
\begin{equation}
    m_B = M_b + \mu_{\rm th}(z) ,
\end{equation}
where $M_b$ denotes the absolute magnitude of SNIa.

The $\chi^2$ function for the supernova-only subsample reads
\begin{equation}
    \chi^2_{\rm SN} = 
    \Delta m_B^{\rm T} \, C_{\rm SN}^{-1} \, \Delta m_B ,
\end{equation}
with
\begin{equation}
    \Delta m_B = m_B^{\rm obs} - M_b - \mu_{\rm th}(z) ,
\end{equation}
where $C_{\rm SN}$ denotes the full covariance matrix including 
both statistical and systematic uncertainties. 

For the Cepheid-calibrated subsample, the likelihood is built from
\begin{equation}
    \chi^2_{\rm Ceph} = 
    \Delta m_B^{\rm T} \, C_{\rm Ceph}^{-1} \, \Delta m_B ,
\end{equation}
where $\Delta m_B = m_B^{\rm obs} - M_b - \mu_{\rm Ceph}$, and $\mu_{\rm Ceph}$ is the Cepheid distance modulus calibration. 

The total SNIa contribution to the likelihood is then
\begin{equation}
    \chi^2_{\rm SNIa} = \chi^2_{\rm SN} + \chi^2_{\rm Ceph} .
\end{equation}

The full covariance matrix, including both statistical and systematic uncertainties, is used in our analysis. Following the standard treatment, we separate the SNIa-only 
subsample and the Cepheid-calibrated subsample in the 
likelihood, combining them consistently within our MCMC 
framework.

\subsection{Hubble Parameter Measurements $H(z)$}

We also employ 
direct measurements of the Hubble expansion rate $H(z)$ 
from cosmic chronometers over
the redshift interval $0.07< z < 1.965$ \citep{Farooq:2017}. Although these measurements exhibit relatively large uncertainties, they are obtained in a
fully model-independent manner using the cosmic
chronometer approach.
These measurements provide an important consistency 
test for cosmological models and help constrain the 
expansion history at intermediate redshifts. 

The likelihood 
is computed as a standard $\chi^2$ with respect to these 
data points used:
\begin{equation}
    \chi^2_{H(z)} = 
    \sum_{i} \frac{\left[ H_{\rm th}(z_i) - H_{\rm obs}(z_i) \right]^2}{\sigma_{H,i}^2} ,
\end{equation}
where $H_{\rm th}(z_i)$ is the model prediction, 
$H_{\rm obs}(z_i)$ is the observed value, 
and $\sigma_{H,i}$ is the reported uncertainty of each data point.

\subsection{Big Bang Nucleosynthesis (BBN)}

To further constrain the baryon density, we impose a 
Gaussian prior on the physical baryon density 
$\omega_b = \Omega_b h^2$ from Big Bang Nucleosynthesis (BBN) 
constraints. Following \cite{Schoneberg:2024}, we adopt a mean 
value of $\omega_b = 0.02218$ with an uncertainty of 
$\sigma = 0.00055$. This prior is implemented as an additional $\chi^2$ term in our total likelihood:
\begin{equation}
    \chi^2_{\rm BBN} = 
    \left( \frac{\omega_b - 0.02218}{0.00055} \right)^2 ,
\end{equation}
where $\omega_b = \Omega_b h^2$.

\subsection{NUMERICAL RESULTS}

In this section, we present the numerical results for the cosmological parameters of the NTHDE and $\Lambda$CDM models, obtained using the combined observational datasets described above. Our analysis is based on a full Markov Chain Monte Carlo  exploration of the parameter space, varying the set of parameters
\begin{equation}
\{\Omega^{\rm (cdm)}_0,\ \Omega^{\rm (b)}_0,\ H_0,\ C,\ \delta,\ \gamma,\ M_b\},
\end{equation}
which correspond to the present-day values of the cold dark matter and baryon density parameters, the Hubble constant, the normalization constant of the NTHDE energy density, the Tsallis deformation parameter $\delta = 1-q$, the Rastall parameter $\gamma = \lambda\kappa$, and the absolute magnitude of Type~Ia supernovae, respectively.

We adopt uniform (flat) priors for all free parameters. The total chi--squared function, $\chi^2_{\rm tot}$, is constructed as the sum of contributions from DESI DR2 BAO measurements, Type~Ia supernovae (including Cepheids), direct $H(z)$ observations, and Big Bang Nucleosynthesis (BBN) constraints.

Four independent MCMC chains were run and combined after discarding the initial 25\% as burn-in. Convergence was assessed using the Gelman--Rubin diagnostic, $\mathcal{R}$, and the effective sample size (ESS), computed using the \texttt{GetDist} package. For all parameters, we obtain $\mathcal{R}$ values very close to unity, indicating excellent convergence and statistical reliability of the posterior distributions.

It is well known that a model with a smaller $\chi^2_{\rm min}$ is not necessarily statistically preferred, as it may include poorly constrained parameters. Such models should be penalized for unnecessary complexity. For this reason, in addition to $\chi^2_{\rm min}$, we also evaluate the Bayesian evidence and the Akaike Information Criterion (AIC) to assess model preference.

To estimate the Bayesian evidence $\mathcal{E}$, we apply the Truncated Harmonic Mean Estimator (THME) \cite{Metodiev:2024}, which provides a stable approximation to the marginal likelihood. In this method, likelihood values are written as
\begin{equation}
\mathcal{L} = e^{-\chi^{2}_{\rm tot}},
\end{equation}
and the evidence is computed as a harmonic mean estimator of $1/\mathcal{L}$ after discarding low-likelihood samples via the cutoff
\begin{equation}
\chi^2 < \chi^2_{\rm min} + \Delta,
\end{equation}
where $\Delta$ controls the truncation level. The Bayesian evidence is then approximated by
\begin{equation}
\log \mathcal{E} \approx -\log\left\langle \frac{1}{\mathcal{L}} \right\rangle_{\rm cutoff}.
\end{equation}

Using a truncation parameter of $\Delta = 10$, we obtain
\begin{equation}
\log \mathcal{E}_{\Lambda{\rm CDM}} \simeq -793.22, \qquad
\log \mathcal{E}_{\rm NTHDE} \simeq -795.71,
\end{equation}
and the Bayes factor is
\begin{equation}
\ln K
=
\log \mathcal{E}_{\Lambda{\rm CDM}}
-
\log \mathcal{E}_{\rm NTHDE}
\simeq 2.49,
\end{equation}
indicating weak-to-moderate evidence in favor of the $\Lambda$CDM model according to the Jeffrey scale. Nevertheless, our model remains a statistically viable extension.

We also compute the Akaike Information Criterion (AIC), defined as
\begin{equation}
{\rm AIC} =
\chi^{2}_{\rm min}
+ 2\,M
+ \frac{2\,M(M+1)}{N - M - 1},
\label{eq:AICc}
\end{equation}
where $N$ is the number of data points and $M$ is the number of free parameters. As shown in Table~\ref{tab:bf_compare}, this metric slightly favors $\Lambda$CDM over the model.

\begin{figure}[h!]
    \centering
    \includegraphics[width=\columnwidth]{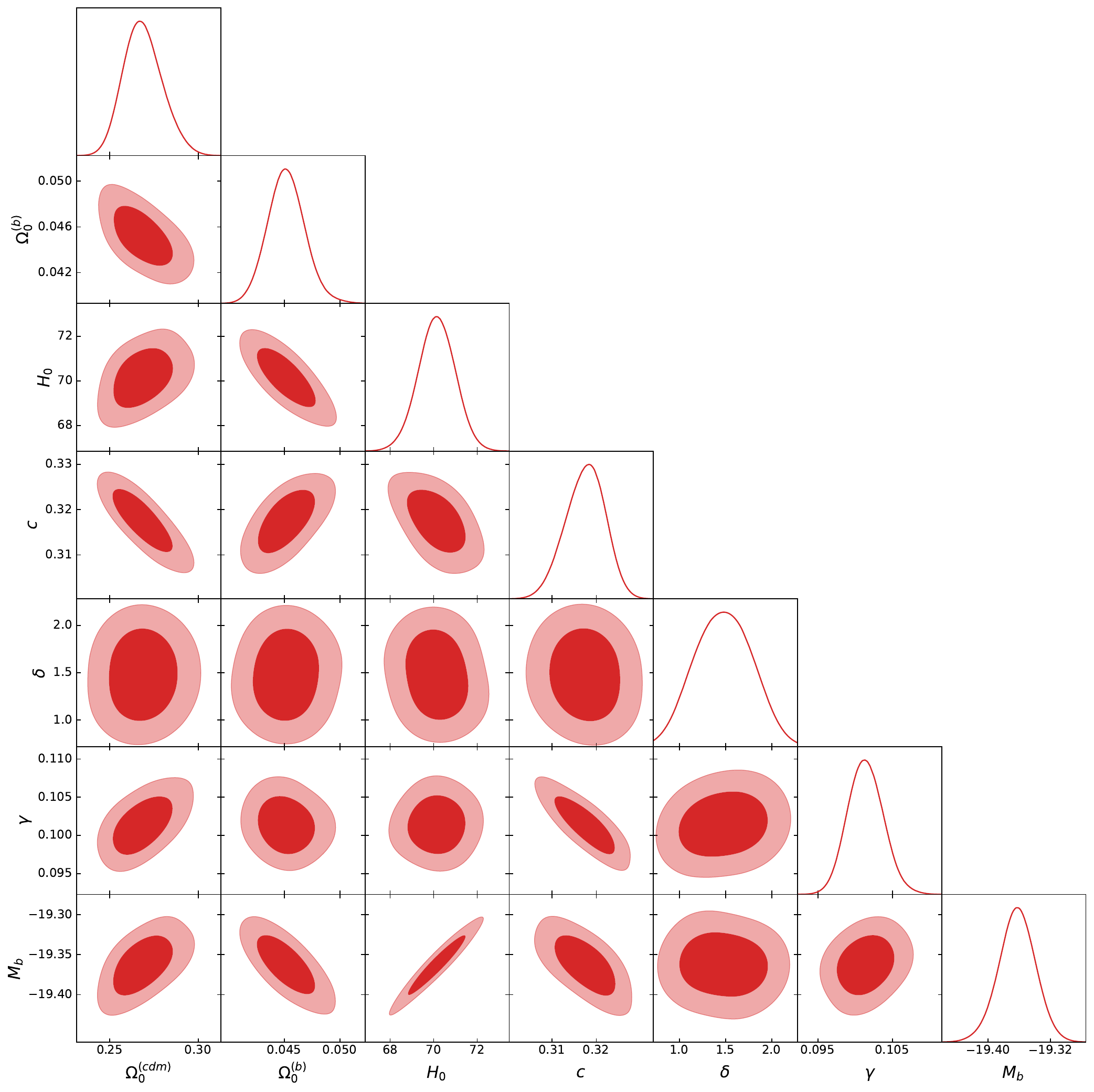}
    \caption{Corner plot showing the marginalized posterior distributions and two-dimensional confidence contours ($68\%$ and $95\%$ CL) for the main cosmological parameters of the NTHDE model, obtained from the combined dataset (SN Ia+Cepheid, $H(z)$, DESI BAO, and BBN).}
    \label{fig:corner_plot_nts}
\end{figure}

For completeness, we also report the reduced chi-squared statistic,
\begin{equation}
\chi^{2}_{\rm red} =
\frac{\chi^{2}_{\rm min}}{N - M},
\label{eq:chi_reduced}
\end{equation}
which quantifies the goodness-of-fit and indicates that both models fit the data comparably well.\\
\begin{table}[h!]
\centering
\caption{Best-fit values of the cosmological parameters for the NTHDE model compared with the $\Lambda$CDM model. Reduced chi-squared ($\chi^2_{\rm red}$), Akaike Information Criterion (AIC), and the Bayesian evidence ($\ln \mathcal{E}$) are also listed.}
\resizebox{\columnwidth}{!}{%
\begin{tabular}{|l|c|c|}
\hline
Parameter & NTHDE (best-fit ) & $\Lambda$CDM (best-fit ) \\
\hline
$\Omega^{ \rm (cdm)}_0$ & $0.270^{+0.010}_{-0.010} $ &  $0.246^{+0.006}_{-0.006}   $ \\
\hline
$\Omega^{ \rm (b)}_0$ & $0.0451^{+0.0014}_{-0.0015}$ & $0.0473^{+0.0007}_{-0.0007}$ \\
\hline
$H_0 \;[\mathrm{km/s/Mpc}]$ & $70.19^{+0.72}_{-0.72}$ & $69.21^{+0.42}_{-0.42} $ \\
\hline
$C$ & $0.316^{+0.004}_{-0.004}   $ & --- \\
\hline
$\delta$ & $1.52^{+0.30}_{-0.30}      $ & --- \\
\hline
$\gamma$ & $0.1018^{+0.0023}_{-0.0023}$ & --- \\
\hline
$M_b$ & $-19.363^{+0.021}_{-0.019} $ & $-19.388^{+0.013}_{-0.013} $ \\
\hline
$\chi^2_{\rm min}$ & $1574.71         $ & $1578.59      $ \\
\hline
$\chi^2_{\rm red} $ & $0.902$ & $0.903$ \\
\hline
$\ln \mathcal{E}$ & $-795.71$ & $-793.22$ \\
\hline
$AIC$  & $1588.71$   &  $1586.59$\\
\hline
\end{tabular}%
}
\label{tab:bf_compare}
\end{table}
\noindent
Figure~\ref{fig:corner_plot_nts} presents the full corner plot of the NTHDE model, showing the one and two-dimensional marginalized posterior distributions of the cosmological parameters obtained from the combined dataset. The diagonal panels show the normalized one-dimensional posteriors, while the off-diagonal panels display the $68\%$ and $95\%$ confidence contours, revealing parameter correlations.
As seen in the figure, the matter-density parameters $\Omega_0^{(\rm cdm)}$ and $\Omega_0^{(b)}$ are only weakly correlated with the model-specific parameters $(C,\delta,\gamma)$, indicating that the introduction of non-extensive and Rastall effects does not significantly distort the standard matter sector. In contrast, a moderate correlation is observed between the Tsallis deformation parameter $\delta$ and the Rastall parameter $\gamma$,  reflecting the coupled role they play in modifying the effective dark-energy dynamics within the NTHDE framework. The parameter $C$, which controls the normalization of the NTHDE energy density, shows a mild degeneracy with $\delta$, consistent with their joint contribution to the late-time expansion rate.
Importantly, the posteriors exhibit well-behaved, nearly Gaussian shapes with compact confidence regions, demonstrating that the MCMC chains have properly converged and that the inferred constraints are statistically robust. No pathological long tails or multimodal structures are observed, confirming that the NTHDE model is stable under joint cosmological observations and provides a consistent fit to the full dataset.

\begin{figure}[h!]
    \centering
    \includegraphics[width=\columnwidth]{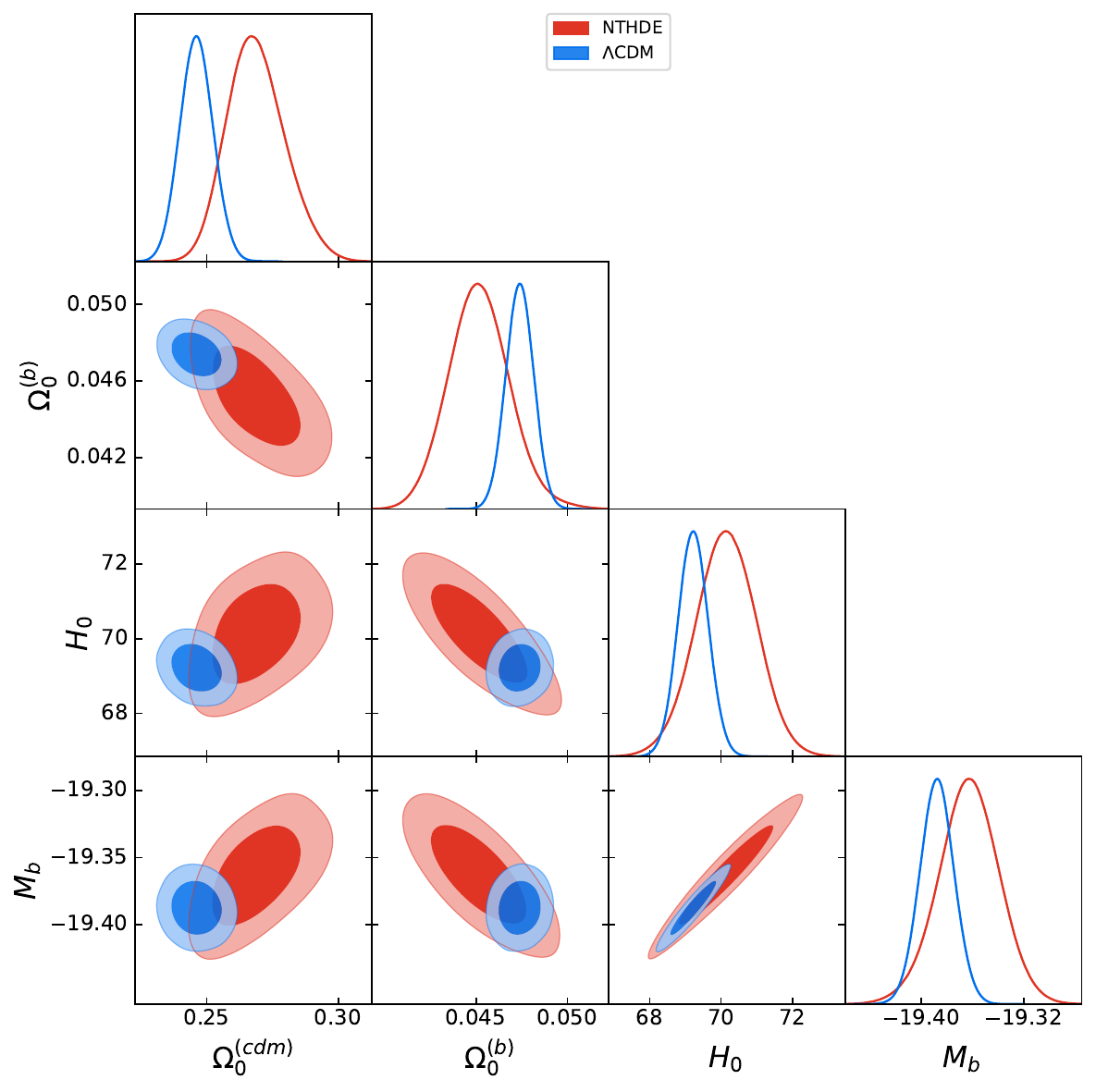}
    \caption{Corner plot showing the marginalized posterior distributions and two-dimensional confidence contours ($68\%$ and $95\%$ CL) for the main cosmological parameters of the NTHDE and $\Lambda$CDM models, obtained from the combined dataset (SN Ia+Cepheid, $H(z)$, DESI BAO, and BBN).}
    \label{fig:corner_plot}
\end{figure}

In addition to analyzing each model separately, we directly compare the posterior distributions of the NTHDE and $\Lambda$CDM models. Figure~\ref{fig:corner_plot} shows the joint marginalized constraints for the key cosmological parameters, where the NTHDE posteriors are plotted together with those of the $\Lambda$CDM$.$ The comparison reveals that both models occupy overlapping regions of parameter space, indicating that the NTHDE scenario can reproduce the expansion history at a level comparable to the standard model. However, noticeable shifts appear in some directions: the NTHDE model prefers slightly higher values of $H_0$ and lower values of $M_b$, as well as a marginally larger effective matter fraction $\Omega_{m}$, compared with $\Lambda$CDM.

\noindent
In addition to the posterior shifts observed in Fig.~\ref{fig:corner_plot}, the parameter 
$\delta$, which quantifies the deviation from standard Boltzmann--Gibbs extensivity through 
$\delta = 1-q$ provides a direct indication of how far the NTHDE model departs from the 
thermodynamic structure underlying the $\Lambda$CDM scenario. The best-fit value 
$\delta \simeq 1.48$, which is significantly larger than unity, indicates a strong non-extensive 
contribution to the effective dark energy sector. Since the limit $\delta = 0$ corresponds to 
the extensive case that reproduces $\Lambda$CDM, the fact that $\delta > 1$ implies that the 
NTHDE cosmology operates in a regime where the modified entropy and informational content of 
spacetime play a non-negligible role. This substantial departure from extensivity is consistent 
with the broader displacement of the posterior contours between the two models and reflects a genuinely different dynamical behaviour, rather than a small perturbation around the 
$\Lambda$CDM limit.

\noindent
From the combined dataset, we also extract several derived cosmological quantities.
The cosmic age is found to be consistent with the Planck $\Lambda$CDM measurements,
while the present-day deceleration parameter $q_0 \approx -0.62$ confirms the accelerated expansion of the Universe.
A detailed comparison of the derived density parameters, including 
$\Omega^{ (m)}_0$, $\Omega^{ (r)}_0$, $\Omega^{ (v)}_0$, the Rastall contribution $\Omega^{(R)}_0$,
and the dark-energy fraction $\Omega^{ (D)}_0$, is presented in Table~\ref{tab:derived_params}.
These quantities highlight the phenomenological differences between NTHDE and $\Lambda$CDM 
at the background level.

\begin{table}[h!]
\centering
\caption{Comparison of derived cosmological parameters for NTHDE and $\Lambda$CDM.}
\begin{tabular}{|l|c|c|}
\hline
{Quantity} & {NTHDE} & {$\Lambda$CDM} \\
\hline
$t_0$ [Gyr]            & 13.90 & 13.68 \\ \hline
$q_0$                  & -0.62 & -0.56 \\ \hline
$\Omega^{ (m)}_0$       & 0.31  & 0.28 \\ \hline
$\Omega^{ (r)}_0$       & $0.84\times10^{-4}$ &  $0.85\times10^{-4}$\\ \hline
$\Omega^{ (v)}_0$     & 0.81 & --- \\ \hline
$\Omega^{(R)}_0$       & 0.12 & --- \\ \hline
$\Omega^{ (D)}_0$       & 0.69  & 0.71\\
\hline
\end{tabular}
\label{tab:derived_params}
\end{table}

\noindent
Here, unlike the standard $\Lambda$CDM scenario in which cosmic acceleration is sourced by a single component (the cosmological constant), the model exhibits a two–component acceleration mechanism. In particular, both the holographic energy sector and the Rastall contribution act as independent accelerating sources, together dominating the late–time dynamics. This split of the total dark sector into two comparable accelerating contributions is a distinctive physical feature of the current model.

\noindent
Until now, we have provided a background-level analysis of the model, including the cosmic expansion history and derived parameters. Detailed plots of $H(z)$, deceleration parameter $q(z)$, $\mu(z)$, $\Omega_m(a)$, and $\Omega_D(a)$ are produced and discussed at the following.

\noindent
In the following, we present the numerical results in more detail by plotting key cosmological quantities as functions of redshift and comparing the NTHDE model's predictions with those of the standard $\Lambda$CDM scenario.

\begin{figure}[h!]
    \centering
    \includegraphics[width=\columnwidth]{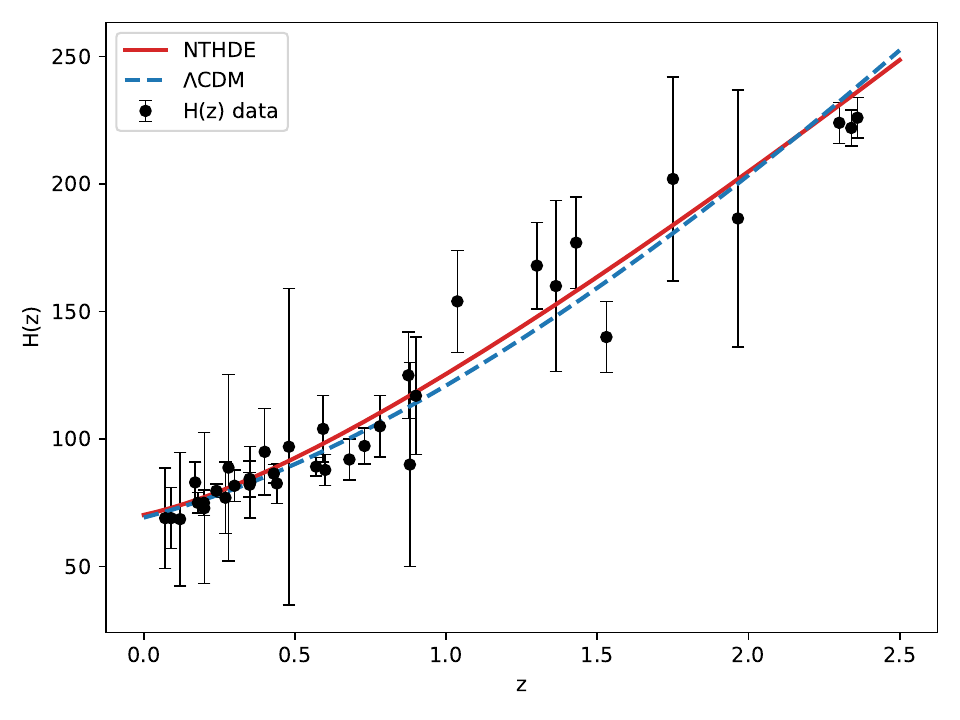}
    \caption{Hubble parameter $H(z)$ as a function of redshift for the NTHDE (solid line) and $\Lambda$CDM (dashed line) models, compared with observational data points.}
    \label{fig:Hz}
\end{figure}

Figure~\ref{fig:Hz} illustrates the evolution of the Hubble parameter $H(z)$ for both the NTHDE and $\Lambda$CDM models. At low redshifts, the two models display nearly identical behavior and closely follow the observational $H(z)$ measurements. However, at higher redshifts ($z \sim 2.5$), the $\Lambda$CDM prediction (dashed curve) begins to deviate from the data and tends to overestimate the expansion rate. In contrast, the NTHDE model (solid curve) remains much closer to the observational points and provides a visibly better fit. This improved agreement at high redshift demonstrates the NTHDE's enhanced ability to reproduce the Universe's expansion history compared to the standard $\Lambda$CDM model.

\begin{figure}[h!]
    \centering
    \includegraphics[width=\columnwidth]{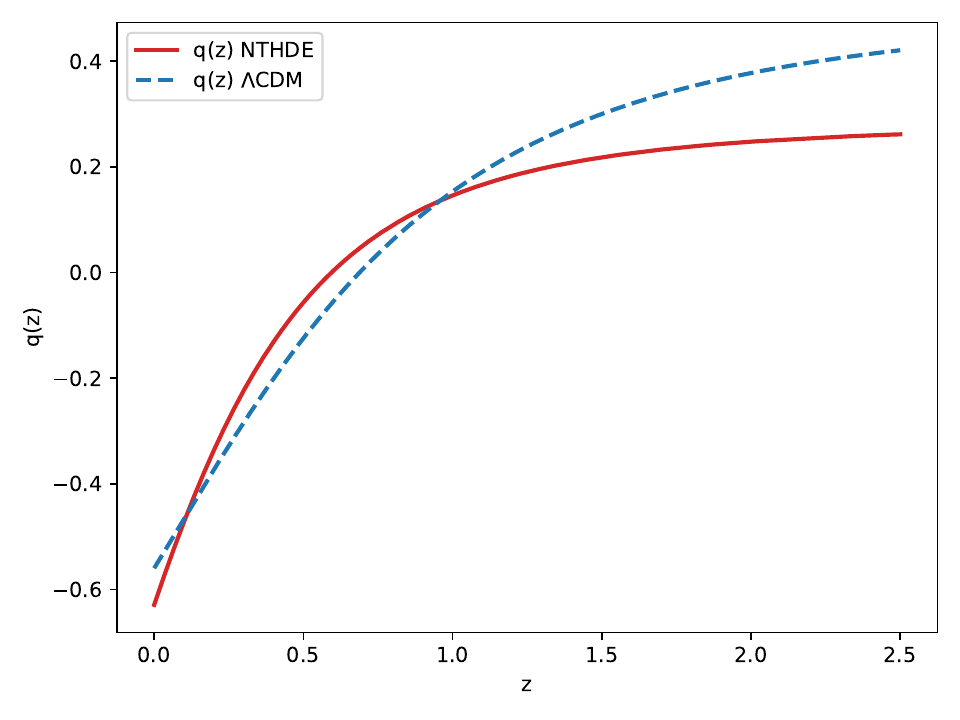}
    \caption{Deceleration parameter $q(z)$ for NTHDE and $\Lambda$CDM models as a function of redshift.}
    \label{fig:deceleration}
\end{figure}

\noindent
Next, Figure~\ref{fig:deceleration} displays the evolution of the deceleration parameter $q(z)$ for both NTHDE and $\Lambda$CDM models. The two curves show distinct behavior. In the present epoch ($z\simeq 0$), the NTHDE model predicts a significantly more negative value of $q(0)$, indicating a stronger rate of accelerated expansion compared to $\Lambda$CDM. Moreover, the transition from deceleration to acceleration occurs earlier in the NTHDE scenario, around $z\simeq 0.5$, whereas the $\Lambda$CDM model undergoes this transition at a slightly higher redshift, $z\simeq 0.6$. 

At intermediate and high redshifts ($z \gtrsim 1$), the difference between the two models becomes more pronounced: the NTHDE model yields systematically smaller values of $q(z)$, reflecting a more rapidly evolving dark-energy component. At the same time, $\Lambda$CDM exhibits a smoother and more gradual evolution. The marked difference in the shapes of the two curves highlights the non-standard and dynamically richer behavior of cosmic acceleration in the NTHDE framework relative to the conventional $\Lambda$CDM model.

\begin{figure}[h!]
    \centering
    \includegraphics[width=\columnwidth]{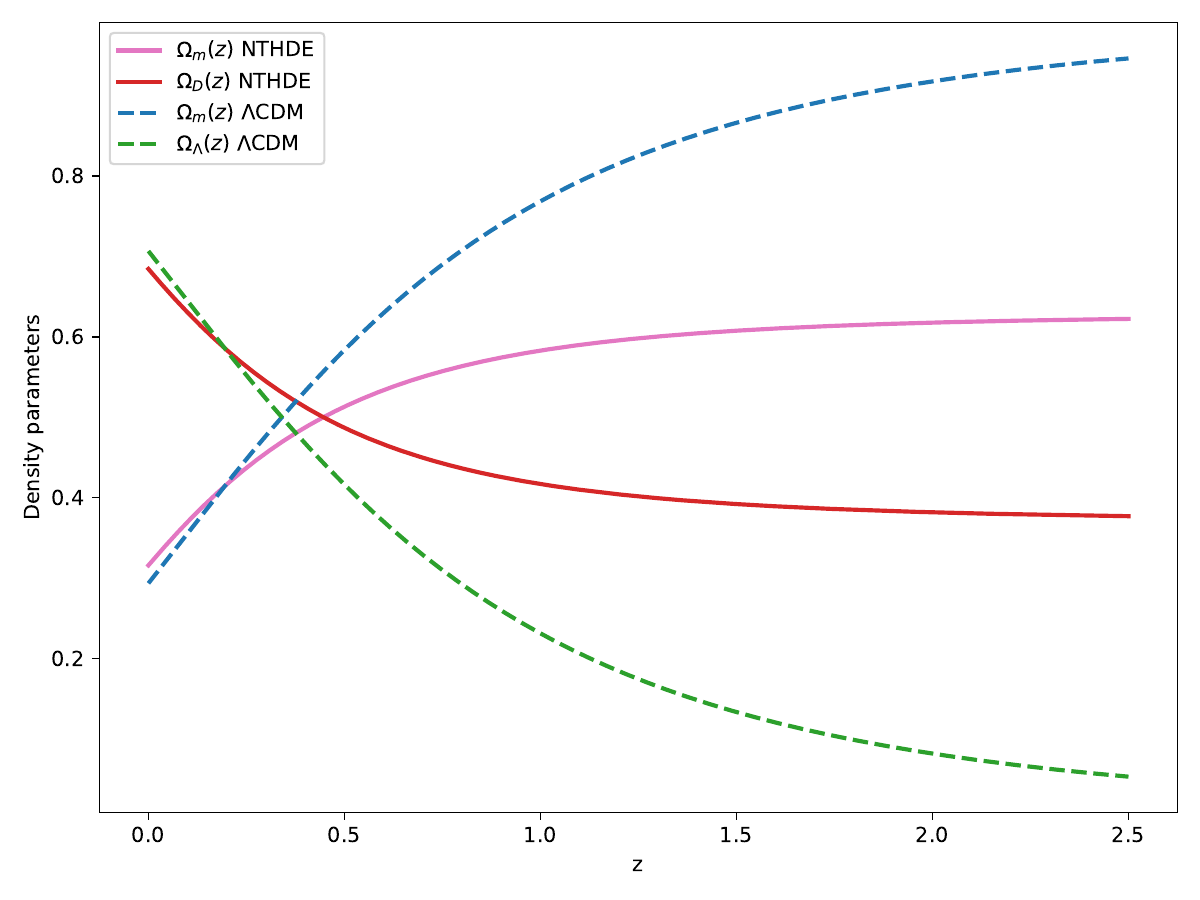}
    \caption{Matter and dark energy density parameters as functions of redshift for NTHDE (solid lines) and $\Lambda$CDM (dashed lines).}
    \label{fig:Omega}
\end{figure}

\noindent
Figure~\ref{fig:Omega} presents the evolution of the matter density parameter $\Omega_m(z)$ and the dark-energy density parameter $\Omega_D(z)$ for both the 
NTHDE and $\Lambda$CDM cosmological models. The two models exhibit markedly different behaviors across the entire redshift range. In the NTHDE model, $\Omega_m(z)$ decreases more slowly with redshift. In contrast, $\Omega_D(z)$ remains 
significantly larger at intermediate and high redshifts, indicating a dynamical dark-energy component that does not dilute as rapidly as in $\Lambda$CDM. 
In contrast, the $\Lambda$CDM model shows a much steeper decline in $\Omega_m(z)$ and a correspondingly rapid drop of $\Omega_\Lambda(z)$ toward the past.

These differences lead to a substantially modified matter–dark-energy balance in the NTHDE scenario at earlier epochs, whereas both models converge to similar present-day values, $\Omega_m(z=0) \simeq 0.3$ and $\Omega_D(z=0) \simeq 0.7$. The distinct shape of the NTHDE curves reflects the non-standard evolution predicted by the Tsallis holographic dark-energy framework.

\begin{figure}[h!]
    \centering
    \includegraphics[width=\columnwidth]{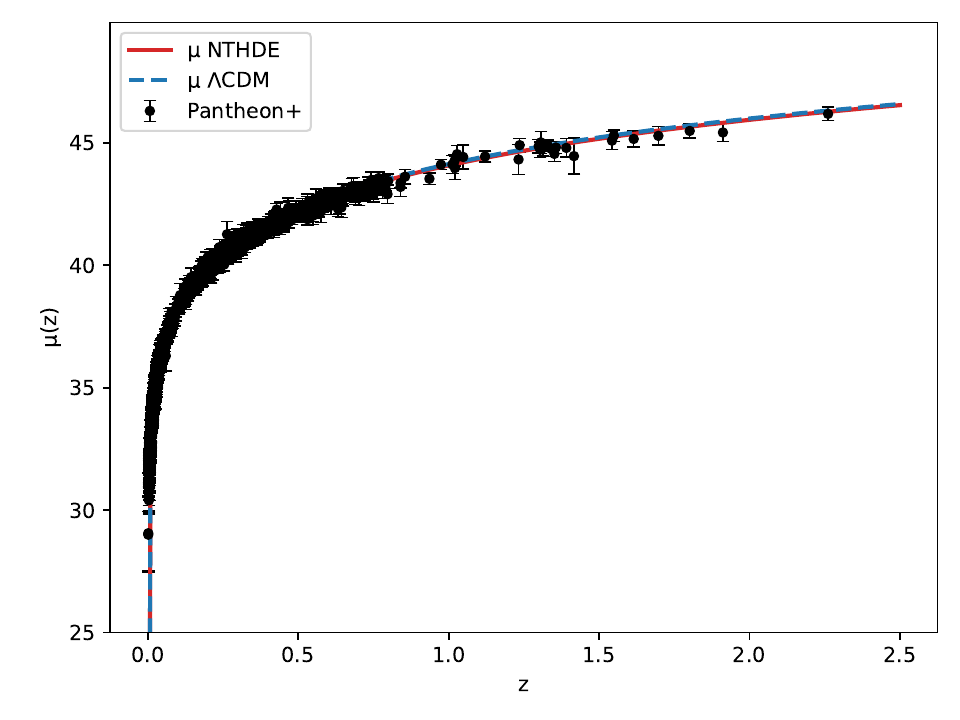}
    \caption{Distance modulus $\mu(z)$ as a function of redshift for NTHDE (solid line) and $\Lambda$CDM (dashed line), compared to Pantheon+ data points.}
    \label{fig:mu}
\end{figure}

\noindent
Figure~\ref{fig:mu} shows the evolution of the distance modulus $\mu(z)$ as a function of redshift, compared with the Pantheon+ compilation of Type Ia supernova data. The theoretical predictions from both the NTHDE and $\Lambda$CDM models are plotted for comparison. Both models provide an excellent fit to the observational data across the entire redshift range ($0 < z < 2.5$), indicating that the NTHDE scenario is fully consistent with current cosmological observations of the late-time Universe.

At low redshifts ($z \lesssim 0.5$), the curves corresponding to the NTHDE and $\Lambda$CDM models nearly overlap, demonstrating that the NTHDE framework reproduces the same luminosity–distance relation as the standard cosmological model in the nearby Universe. As redshift increases, a slight deviation appears, but it remains well within the observational uncertainties of the Pantheon+ data set. This agreement suggests that the NTHDE model not only retains the predictive success of $\Lambda$CDM but also provides a more general phenomenological description that allows for dynamical dark energy behavior at higher redshifts.

\noindent
In summary, the NTHDE model reproduces the cosmic expansion history, the deceleration parameter, the density evolution of matter and dark energy, and the distance modulus in good agreement with observational data. Compared with $\Lambda$CDM, small but measurable deviations appear at intermediate redshifts, which could serve as potential observational signatures to distinguish NTHDE from the standard cosmological scenario.









\section{Summary and concluding remarks}

In this work, we have investigated the cosmological implications of the New Tsallis Holographic Dark Energy (NTHDE) within the framework of Rastall gravity and confronted the model with the latest background-level observational data. Using the corresponding modified Friedmann equations and performing a full Markov Chain Monte Carlo analysis, we obtained robust constraints on the main cosmological parameters, including the Tsallis deformation parameter $\delta$ and the Rastall parameter $\gamma$, both of which characterize deviations from standard statistical mechanics and general relativity, respectively. As a significant difference from previous studies on NTHDE \cite{Moradpour:2020dfm, Pandey:2021fvr, Waheed:2023flg}, here, a combination of the Rastall term and NTHDE, called $\Omega_D$, seems able to mimic the current value of the dark energy density parameter predicted in the $\Lambda$CDM model.

The model reproduces the late-time accelerated expansion and yields cosmological quantities such as the Hubble parameter $H_0$, the universe age, the present-day deceleration parameter $q_0$, and the density parameters $\Omega_m$, and $\Omega^r$ in agreement with the current observational expectations. Indeed, as it is obvious from Tables~\ref{tab:bf_compare} and \ref{tab:derived_params}, $\Omega_D$ is prone to mimic the current value of the density parameter reported by the $\Lambda$CDM model. Moreover, if one looks at $\Omega_\nu$ as the sole dark energy, then all info of Tables~\ref{tab:bf_compare} and \ref{tab:derived_params} are still available. In this manner, deviation from the prediction of the $\Lambda$CDM model about the current value of density parameter of dark energy would be impressive. Taken together, these results indicate that NTHDE in Rastall framework provides a viable alternative to $\Lambda$CDM at the background level, capable of describing the cosmic expansion history while naturally incorporating generalized statistical modifications.

By employing the Pantheon+SH0ES Type Ia supernova sample, the DESI DR2 BAO measurements, cosmic chronometer $H(z)$ data, and the Big Bang Nucleosynthesis prior on $\omega_b$, we showed that the NTHDE model provides an excellent fit to the current expansion-history observations. The constrained value of the Rastall parameter ($\gamma$) remains fully compatible with previous reports addressing the permissibly of $\gamma<1/6$ and $\gamma>1/4$ \cite{Moradpour:2016fur, Moradpour:2017ycq, Li:2019jkv, ElHanafy:2022kjl, Afshar:2023uyw}. Therefore, it is indicated that only mild departures from standard energy–momentum conservation are needed. Moreover, the Tsallis deformation parameter is found to be significantly larger than zero, suggesting that non-extensive entropy effects may play a non-negligible role in the effective dark-energy dynamics.

\section{DATA AVAILABILITY}

The data used in this work are available publicly.

\section{acknowledgment}
ZD is supported by the Korea Institute for Advanced Study (KIAS) under grant no.6G097301.






\end{document}